\documentclass[12pt]{article}
\usepackage{epsfig}
\usepackage{graphics}
\usepackage{psfrag}
\usepackage{amsmath}
\usepackage{amssymb}
\usepackage{color}

\newcommand{\bea}{\begin{eqnarray}}
\newcommand{\eea}{\end{eqnarray}}
\newcommand{\simgt}{\hbox{ \raise3pt\hbox to 0pt{$>$}\raise-3pt\hbox{$\sim$} }}
\newcommand{\simlt}{\hbox{ \raise3pt\hbox to 0pt{$<$}\raise-3pt\hbox{$\sim$} }}

\newcommand{\LQ}{\Lambda_{\rm QCD}}

\begin{document}

\begin{titlepage}

    \begin{flushright}
      \normalsize UT--15--38 \\
      \normalsize TU--1006\\
      \today
    \end{flushright}

\vskip2.5cm
\begin{center}
\large\bf\boldmath
Determination of $m_c$ and $m_b$ from
quarkonium\\ 
$1S$
energy levels
in perturbative QCD
\vspace{10mm}\\
\unboldmath
\end{center}

\vspace*{0.8cm}
\begin{center}
{\sc Y. Kiyo}$^{a}$,
{\sc G. Mishima}$^{b}$ and
{\sc Y. Sumino$^{c}$}\\[5mm]
  {\small\it $^a$ Department of Physics, Juntendo University}\\[0.1cm]
  {\small\it Inzai, Chiba 270-1695, Japan}

  {\small\it $^b$ Department of Physics, University of Tokyo}\\[0.1cm]
  {\small\it  Bunkyo-ku, Tokyo 113-0033, Japan}

  {\small\it $^c$ Department of Physics, Tohoku University}\\[0.1cm]
  {\small\it Sendai, 980-8578 Japan}

\end{center}

\vspace*{2.8cm}
\begin{abstract}
\noindent
We update determination of the
$\overline{\rm MS}$ masses of the charm and bottom quarks,
from comparisons of the masses of the
charmonium and bottomonium $1S$ states with their perturbative
predictions up to next-to-next-to-next-to-leading order
in $\varepsilon$ expansion and using the $\overline{\rm MS}$ masses.
Effects of non-zero charm-quark mass in the bottomonium masses
are incorporated up to next-to-next-to-leading order.
We obtain 
$
\overline m_c=1246\pm 2 (d_3) \pm 4 (\alpha_s) \pm 23 (\text{h.o.} )~{\rm MeV}
$
and
$
\overline m_b=4197\pm 2 (d_3) \pm 6 (\alpha_s) \pm 20 (\text{h.o.} )\pm 5  (m_c)~ {\rm MeV}
$,
which agree with the current Particle Data Group values.
\vspace*{0.8cm}
\noindent

\end{abstract}


\vfil
\end{titlepage}

\newpage

The Standard Model (SM) of elementary particle physics
has been successfully completed by the discovery
of the Higgs particle, whereas no definite clue to 
physics beyond the SM has been found yet.
In such an era, necessity for precise understanding of
the SM is increasing more than ever.
In particular, to meet demands
for accurate measurements required in the LHC experiments
as well as those for high-precision 
flavor physics, etc., there has been remarkable
progress in the predictability of
perturbative QCD in recent years. 
The masses of the  $c$- and $b$-quarks are among
the important fundamental parameters of perturbative QCD.
They play crucial roles, for instance,
in testing predictions of the SM for the Yukawa 
coupling constants of the Higgs particle,
and also as the input parameters for predicting various
observables in flavor physics.

The masses of the $c$- and $b$-quarks have been 
determined in many ways.
Even referring only to
works published after the latest version of
Review of Particle Physics \cite{Beringer:1900zz} by Particle Data Group (PDG) Collaboration,
there are analyses based on
non-relativistic QCD sum rule \cite{Penin:2014zaa,Beneke:2014pta},
relativistic QCD sum rule \cite{Dehnadi:2015fra},
deep inelastic scattering \cite{Abramowicz:2014zub},
heavy quarkonium spectroscopy \cite{Ayala:2014yxa},
and lattice computation \cite{Yang:2014sea}.
(See \cite{Beringer:1900zz} for earlier studies.)
Their physics ingredients vary substantially, and also they 
probe different kinematical regions of QCD.
Therefore, consistency of the determined values provides
a non-trivial test of QCD, and of the SM more generally.

In this paper we determine the masses of the $c$- and $b$-quarks
from comparisons of the energy levels of the
charmonium and bottomonium $1S$ states with their perturbative
predictions.
This is an update of the mass determination
performed as part of the analyses in 
\cite{Brambilla:2001fw,Brambilla:2001qk}, which included
perturbative expansion 
up to the next-to-next-to-leading order (NNLO).
We include one more order, namely up to NNNLO.
Recently, the four-loop relation to the pole 
and $\overline{\rm MS}$ quark masses has been
computed \cite{Marquard:2015qpa}.
The present study is the first full analysis using
the $\overline{\rm MS}$ mass up to NNNLO.
We also include non-zero charm-quark mass effects
in the computation of the bottomonium energy levels,
up to the highest order of the currently available computations
(up to two loops of internal $c$-quark) \cite{Hoang:2000fm}.
On the experimental side, accurate data on the
$\eta_c(1S)$ and $\eta_b(1S)$ masses are available today,
which we include in our analysis, in addition to
the $J/\psi(1S)$ and $\Upsilon(1S)$ masses
used in the NNLO analyses.

The purpose of the present study is to provide another
accurate determination of these quark masses, and also
there is a different aspect.
The heavy quarkonium states are unique among various hadrons, 
in that properties of individual hadronic states
can be predicted purely within perturbative QCD.
Hence, if we observe consistency with the masses determined by 
other methods with high accuracy,
that can be an evidence that pure
perturbative QCD is indeed capable of predicting
properties of these individual hadrons with high precision,
with only $\alpha_s$ and the quark masses as the
input parameters of the theory.

Since the study at the previous order,
our understanding based on perturbative QCD has developed
considerably.
On the one hand, developments in computational technology
enabled (finally) accomplishment of the full NNNLO computation
of the quarkonium energy levels, which have been
carried out stepwise.
Milestone computations include computations of 
the two-loop $1/(mr^2)$ potential \cite{Kniehl:2001ju},
the full NNNLO Hamiltonian \cite{Kniehl:2002br},
the three-loop static potential \cite{Anzai:2009tm},
full formula of the spectrum \cite{Kiyo:2014uca},
the four-loop relation between the pole and 
$\overline{\rm MS}$ quark masses \cite{Marquard:2015qpa}, etc.
In addition, non-zero quark mass effects
in the three-loop pole-$\overline{\rm MS}$
mass relation have been computed \cite{Bekavac:2007tk}.

On the other hand, deeper understanding on the structure
of QCD has been achieved in the meantime.
Solid theoretical backgrounds have been formed
based on effective field theory (EFT) frameworks,
such as potential-NRQCD \cite{Pineda:1997bj} and velocity-NRQCD \cite{Luke:1999kz}.
Accumulation of empirical facts also reinforced our
understanding.
There were examinations of various higher-order perturbative 
predictions
and computations by lattice QCD simulations.
New experimental data on heavy quarkonium states,
such as $\eta_c$, $\eta_b$ and other $c\bar{c}$, $b\bar{b}$ 
states, 
became available.
Detailed comparisons of these results clarified the
status of perturbative QCD predictions in an
unequivocal manner 
\cite{Sumino:2005cq,Anzai:2009tm,Bauer:2011ws,Marquard:2015qpa}.
For instance, relations between renormalons and 
non-perturbative matrix elements in EFT became clearer,
which are supported by growing number of evidences.
In purely perturbative predictions, infrared
(IR) contributions are encoded as IR renormalons, which induces
an uncertainty of the order of non-perturbative matrix
elements \cite{Sumino:2014qpa}.
By contrast, in an operator product expansion (OPE) in EFT, 
one should subtract IR renormalons from perturbative evaluations
of Wilson coefficients and replace the renormalons by
non-perturbative matrix elements.
(See, e.g., discussion on estimates of non-perturbative
contributions
to the heavy quarkonium energy levels in \cite{Kiyo:2013aea}.)

Since the analysis \cite{Ayala:2014yxa} uses a method similar to
ours for determination of the $b$-quark mass, we state the differences
of our study.
(1)We include the exact four-loop pole-$\overline{\rm MS}$
mass relation \cite{Marquard:2015qpa},
which became available only after the study \cite{Ayala:2014yxa}.
(2)We include non-zero charm mass effects on the bottomonium energy levels 
up to two loops of the $c$-quark \cite{Hoang:2000fm}, 
whereas \cite{Ayala:2014yxa}
includes only up to one loop.
(3)Ref.\,\cite{Ayala:2014yxa} uses the renormalon-subtracted mass 
at intermediate stage, whereas we compute the energy levels
directly in terms of the $\overline{\rm MS}$ mass.
(4)We use the strong coupling constant of four active quark flavors 
in the reference  
analysis of the bottomonium energy levels, in contrast to the
three-flavor coupling used in \cite{Ayala:2014yxa}.

\subsection*{Determination of $c$-quark mass from $J/\psi(1S)$ and $\eta_c(1S)$}

In perturbative QCD,
the energy level of a charmonium state is given by
\begin{align}
M_{c\bar c}=2m_c^{\rm pole}+E_{\text{bin},\,c\bar{c}} \, .
\label{tot}
\end{align}
The pole mass of the $c$-quark is expressed in
terms of the $\overline{\rm MS}$ 
mass as
\begin{align}
m^{\rm pole}=\overline m\left[ 1+ \sum _{k=0}^3 d_k \left( \frac{ \varepsilon \,
\alpha _s(\overline m)}{\pi} \right)^{k+1} +\mathcal{O}( \varepsilon ^5)\right],
\label{mass}
\end{align}
where $\overline m \equiv m^{\overline{\rm MS}}(m^{\overline{\rm MS}})$ 
represents the $\overline{\rm MS}$ mass renormalized at the
$\overline{\rm MS}$ mass.
We use the $\varepsilon$--expansion \cite{Hoang:1998ng} to cancel the 
${\cal O}(\LQ)$ renormalons in $2m^{\rm pole}$ and $E_{\rm bin}$.
In the computation of the charmonium levels, we use the coupling constant
with $n_f=3$ active quark flavors,
$\alpha_s^{(3)}$.
The values of $d_k$ are taken from \cite{Chetyrkin:1999qi,Melnikov:2000qh,Marquard:2015qpa}, which
are converted from their values in the theory 
with 4 flavors (with $c$-quark)
to those with 3 flavors (without $c$-quark), 
using the matching relation \cite{Chetyrkin:1997sg}
\begin{align}
\alpha ^{(n_f+1)} _s(\overline m)=
\alpha ^{(n_f)} _s(\overline m)
\left[ 1-\frac{11}{72} 
\frac{
\alpha ^{(n_f)} _s(\overline m)^2}
{\pi^2}
-\left( \frac{564731}{124 416}-\frac{82043}{27 648}\zeta (3)-\frac{2633}{31 104}n_f\right) 
\frac{
\alpha ^{(n_f)} _s(\overline m)^3}
{\pi^3}
\right] .
\label{match}
\end{align}
We obtain
\begin{align}
d_0=4/3, ~~
d_1=10.3193, ~~
d_2=116.300, ~~
d_3=1687.1\pm 21.5 \, .
\label{dn}
\end{align}
Then we express $\alpha_s(\overline m)$ by the series expansion
in $\alpha_s(\mu)$ using
the relation
\begin{align}
\alpha_s(\overline m) = \alpha_s(\mu)\Biggl[
1 &+ \frac{\beta_0}{2}\log\biggl(\frac{\mu}{\overline m}\biggr)
\left( \frac{ \varepsilon \alpha_s(\mu)}{\pi} \right)
\nonumber \\ &
 + \Biggl(\frac{\beta_0^2}{4}\log^2\biggl(\frac{\mu}{\overline m}\biggr)
 + \frac{\beta_1}{2}\log\biggl(\frac{\mu}{\overline m}\biggr) \Biggr)
\left( \frac{ \varepsilon \alpha_s(\mu)}{\pi} \right)^2
+ \dots
\Biggr]\, ,
\end{align}
which follows from the renormalization-group (RG)
equation for $\alpha_s$;
$\beta_i$ denotes the $(i+1)$-loop coefficient of the
beta function.
[We suppressed the ${\cal O}(\varepsilon^3)$ term for brevity.]
Here and hereafter, $\alpha_s$ represents
$\alpha_s^{(n_f)}$.

The binding energy is given by \cite{Beneke:2005hg,Penin:2002zv,Kiyo:2014uca}
\begin{align}
E_{\rm bin}=-\frac{4}{9} \alpha_s(\mu)^2 m^{\rm pole} \sum _{k=0}^3 \varepsilon ^{k+1} \left(\frac{\alpha_s(\mu)}{\pi}\right)^k P_k(L_\mu)
+\mathcal{O}(\varepsilon ^5)\, .
\label{bin}
\end{align}
Here, $P_k(L_\mu)$ is a $k$-th order polynomial of
$L_\mu =\log [3\mu /(4 \alpha_s (\mu) m^{\rm pole} )]+1$.
Apart from $c_k\equiv P_k(0)$,
the polynomial is determined by the RG equation for
$\alpha_s$.
$c_k$'s are given, for $J/\psi(1S)$ $[n_f=3, n=1, l=0, s=1, j=1]$, by
\begin{align}
c_1 = 7/2,~~
c_2 = 142.018~~
c_3=1276.83(1) + 474.289 \log \alpha_s(\mu)\, ,
\end{align}
and for $\eta_c (1S)$ $[n_f=3, n=1, l=0, s=0, j=0]$, by
\begin{align}
c_1 = 7/2,~~
c_2 = 165.413,~~
c_3= 908.82(1) + 597.111 \log \alpha_s(\mu)\, .
\end{align}

\begin{figure}[t]
\begin{center}
\includegraphics[width=8cm]{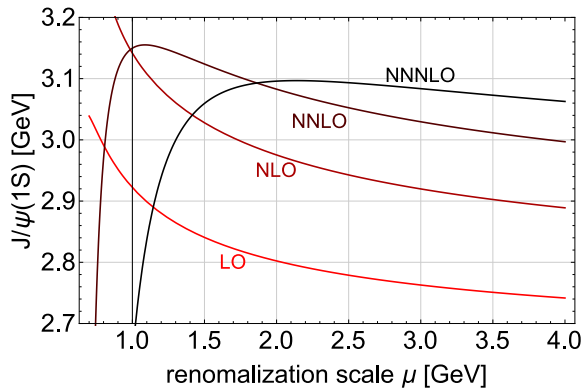}
\end{center}
\caption{\small
Scale dependences of the predictions for the $J/\psi(1S)$ mass.
N$^k$LO line represents the prediction up to ${\cal O}(\varepsilon^{k+1})$.
We take
$\overline m_c=1266$~MeV, which is adjusted to
reproduce the experimental data at the
minimal sensitivity scale for the NNNLO line.
}
\label{fig:charmonium-scale-dep}
\end{figure}

The input value for $\alpha_s$ is set as \cite{Beringer:1900zz}
\begin{align}
\alpha_s ^{(n_f=5)}(m_Z)=0.1185\pm 0.0006\, .
\label{alpha}
\end{align}
Evolving by the RG equation,
it is matched to the couplings with 4 and 
3 flavors successively,
using the matching relation eq.~(\ref{match}).
We compare the predictions of the $J/\psi(1S)$ and $\eta_c(1S)$
masses with the experimental data \cite{Beringer:1900zz}:
\begin{align}
M_{J/\psi(1S)}^\text{exp}=3096.916\pm 0.011 ~~ {\rm MeV},
~~~~~
M_{\eta_c(1S)}^\text{exp}=2983.6\pm 0.7 ~~ {\rm MeV}\, .
\end{align}
The scale dependences of the predictions are shown
in Fig.~\ref{fig:charmonium-scale-dep}
for $J/\psi(1S)$.
Those for $\eta_c(1S)$ are  similar.
The scale dependences decrease as the order is raised.
Fig.~\ref{fig:charmonium-scale-dep} is also
consistent with the expectation of
renormalon dominance that
the minimal sensitivity scale increases with the order \cite{Sumino:2014qpa}.
We can adjust the value of $\overline m_c$
to reproduce the experimental data at the minimal sensitivity scales.
The central values read
$\overline m_c=1266$~MeV and 1226~MeV, respectively, 
for $J/\psi(1S)$ and $\eta_c(1S)$, and
the minimal sensitivity scales for the predictions up to NNNLO read
$\mu =2.14$~GeV and 2.42~GeV, respectively.
(The values of $\alpha_s(\mu)$ are 0.2712
and 0.2878, respectively.)
The $\varepsilon$--expansions at the minimal sensitivity scales
are given by
\begin{align}
M_{J/\psi(1S)}^\text{pert}&=2532 + 263 + 170 + 109+ 23 ~~ {\rm MeV}\, ,
\label{predicJpsi}\\
M_{\eta_c(1S)}^\text{pert}&=2452 + 242 + 162 + 103 + 24~~ {\rm MeV}\, ,
\label{predicetac}
\end{align}
where the terms on the right-hand side
correspond to the order $\varepsilon^0$,
$\varepsilon^1$, $\dots$, $\varepsilon^4$ terms, respectively. 
They exhibit reasonably convergent behaviors.

We estimate uncertainties of our predictions, which
are translated to uncertainties in the determination
of $\overline m_c$.
(The errors of the experimental data are negligibly small.)
\\
(i) {\it Uncertainty of $d_3$}:
The uncertainty of $d_3$ in eq.~(\ref{dn}) correspond
to $\pm 2$~MeV variation of $\overline m_c$.
Other uncertainties in the parameters in the expansion coefficients,
such as that of $c_3$, are negligibly small.
\\
(ii) {\it Uncertainty of $\alpha_s(M_Z)$}:
The uncertainty in the input $\alpha_s(M_Z)$ in
eq.~(\ref{alpha}) corresponds to $\pm 4$~MeV shift of $\overline m_c$.
\\
(iii) {\it Uncertainty by higher-order corrections}:
We estimate the uncertainty of unknown higher-order corrections
in three different ways.
(a) We change the scale $\mu$ from the minimal sensitivity
scale to twice of that value.
The corresponding variations of $\overline m_c$ are about 18~MeV
and $15$~MeV, respectively,
for $J/\psi(1S)$ and $\eta_c(1S)$.
(b) We take the differences of the determined $\overline m_c$
using the same method up to NNLO and NNNLO, fixing $\mu$
at the respective minimal sensitivity scales.
This results in the differences of 27~MeV and 19~MeV,
respectively, when we adjust $M_{c\bar c}$ to  the $J/\psi(1S)$ and $\eta_c(1S)$
masses.
(The respective minimal sensitivity scales up to NNLO are
1.08~GeV and 1.23~GeV.)
(c) We take one half of the last known terms of the series
in eqs.~(\ref{predicJpsi}) and (\ref{predicetac}).
This results in about 12~MeV for both states.
Let us take the maximal values of (a)--(c) as our estimates of 
higher-order corrections.
They give 27~MeV and 19~MeV for uncertainties of $\overline m_c$ as
determined from the masses of $J/\psi(1S)$ and $\eta_c(1S)$,
respectively.
The corresponding uncertainties for the predictions of
$M_{c\bar c}(1S)$ (twice of these values)
are similar in size to the estimate of uncanceled renormalon
of order $\LQ^3 r^2$, in the case $\LQ\sim 300$~MeV and
$r \sim 1.5$~GeV$^{-1}$ (although it is sensitive
to the values chosen for $\LQ$ and $r$).
As stated, in a purely perturbative prediction, 
this renormalon uncertainty is the substitute for a non-perturbative matrix
element in OPE of EFT.

\begin{figure}[t]
\begin{center}
\includegraphics[width=11cm]{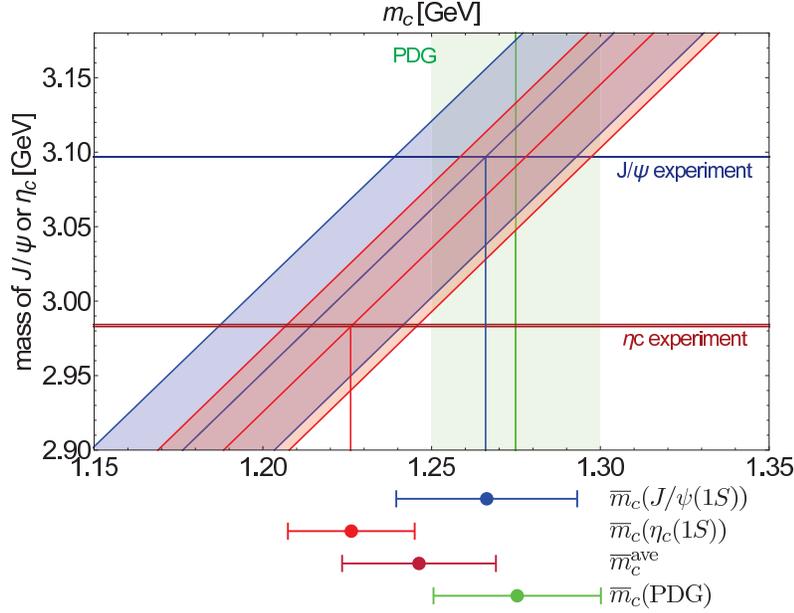}
\end{center}
\vspace*{-5mm}
\caption{\small
Determination of $\overline m_c$.
Horizontal (vertical) axis represents
$\overline m_c$ 
(mass of charmonium $1S$ state).
Horizontal bands denote the experimental data with errors.
Diagonal bands show the perturbative QCD predictions with
errors as functions of $\overline m_c$.
Determined $\overline m_c$ with error bars are shown below the plot.
For comparison, the PDG
value is also shown.
}
\label{fig:charm}
\end{figure}

To summarize our results, we obtain
\begin{align}
\overline m_c({J/\psi(1S)})&=1266\pm 2\ (d_3)\ \pm 4\ (\alpha_s)\ \pm 27\ (\text{h.o.} )\quad {\rm MeV}\, ,
\label{mcbar-Jpsi}\\
\overline m_c({\eta_c(1S)})&=1226\pm 2\ (d_3)\ \pm 4\ (\alpha_s)\ \pm 19\ (\text{h.o.} )\quad {\rm MeV}\, .
\label{mcbar-etac}
\end{align}
Both values are mutually consistent within the estimated errors.
By taking the average of the above two estimates, we obtain
\begin{align}
\overline m_c^\text{ave}&=1246\pm 2\ (d_3)\ \pm 4\ (\alpha_s)\ \pm 23\ (\text{h.o.} )\quad {\rm MeV}\, .
\label{mcbar-ave}
\end{align}
It is consistent with the current PDG
value $\overline m_c=1275\pm25$~MeV \cite{Beringer:1900zz}.
See Fig.~\ref{fig:charm}.

\subsection*{Determination of $b$-quark mass from
$\Upsilon (1S)$ and $\eta_b (1S)$}

In the limit where we neglect masses of quarks
in internal loops,
the formula for the bottomonium energy level is the same
as that for the charmonium, except that we set
$n_f=4$.
It is known, however, that effects of the $c$-quark mass
is important in the predictions of the bottomonium energy levels.
Presently the corrections by non-zero $m_c$ effects
are known up to ${\cal O}(\varepsilon^3)$.
These effects are included in our predictions
in the following way.
\begin{align}
M_{b\bar b}=2m_b^{\rm pole}+E_{\text{bin},\,b\bar{b}}
\label{bb-energy}
\end{align}
with
\begin{align}
&
m_b^{\rm pole}=\Bigl[ m_b^{\rm pole} \Bigr]_{\overline m_c\to 0} +
\overline m_b\left[ 
d_1^{(c)}  \left( \frac{ \varepsilon \, \alpha_s(\overline m)}{\pi} \right)^2
+d_2^{(c)}  \left( \frac{ \varepsilon \, \alpha_s(\overline m)}{\pi} \right)^3
\right] ,
\label{mass_c}
\\
&
E_{\text{bin},\,b\bar{b}}=
\Bigl[ E_{\text{bin},\,b\bar{b}} \Bigr]_{\overline m_c\to 0} +
2m_b^{\rm pole} \Bigl[ 
- \varepsilon ^2 \Delta _{\rm NLO}^{(c)} 
- \varepsilon ^3 \Delta _{\rm NNLO}^{(c)} 
\Bigr]\, .
\end{align}
Here, 
$[ m_b^{\rm pole} ]_{\overline m_c\to 0}$ and
$[ E_{\text{bin},\,b\bar{b}} ]_{\overline m_c\to 0}$,
respectively, represent eqs.~(\ref{mass})
and (\ref{bin}) for $n_f=4$,
and the parameters therein are given by
\begin{align}
d_0=4/3, ~~
d_1=9.27792, ~~
d_2=94.2137, ~~
d_3=1220.3\pm 21.5
~;
\label{d3b}
\end{align}
for $\Upsilon (1S)$ $[n_f=4, n=1, l=0, s=1, j=1]$,
\begin{align}
c_1 = 53/18,~~
c_2 = 125.69,~~
c_3=1010.65(1) + 474.289 \log \alpha_s(\mu)\, ,
\end{align}
and for
$\eta_b (1S)$ $[n_f=4, n=1, l=0, s=0, j=0]$,
\begin{align}
c_1 = 53/18,~~
c_2 = 149.09,~~
c_3=665.70(1) + 597.111 \log \alpha_s(\mu)\, .
\end{align}

The deviation from the limit ${\overline m_c\to 0}$
is parametrized as follows.
At ${\cal O}(\varepsilon^2)$ \cite{Gray:1990yh,Eiras:1999xx},
\begin{align}
d_1^{(c)}(x)=&\frac{1}{18} \Bigl[ \,6 (x+1)^2 \left(x^2-x+1\right) (\text{Li}_2(-x)+\log (x) \log (x+1))
\nonumber\\
&~~~+6 (x-1)^2 \left(x^2+x+1\right) (\text{Li}_2(x)+\log (1-x) \log (x))
\nonumber\\
&~~~-6 x^4 \log ^2(x)-6 x^2 \log (x)-9 x^2-\pi ^2 
\left(x^4-3 x^3-3x\right)\Bigr]\, ,
\label{d1c}
\\
\Delta _{\rm NLO}^{(c)}
=&\frac{4 \alpha_s(\mu)^3}{27 \pi}
\left[
\frac{3\pi}{4}\rho -2\rho^2+\pi \rho^3
+\log \frac{\rho}{2} 
+\frac{2-\rho^2-4\rho^4}{\sqrt{\rho ^2-1}} \mathrm{Arctan}
\biggl(\sqrt{\frac{\rho -1}{\rho +1}}\biggr)
\right] ,
\end{align}
where $x=\overline m_c/\overline m_b$ and 
$\rho =3 m_c^\text{pole}/[2 \alpha_s(\mu) m_b^\text{pole} ]$.
The expressions for $d_2^{(c)}$ and $\Delta _{\rm NNLO}^{(c)}$
are lengthy, which 
can be found in the original references  \cite{Bekavac:2007tk}
and \cite{Hoang:2000fm}, respectively,
and  we refrain from showing them.\footnote{
Concerning $\Delta _{\rm NNLO}^{(c)}$,
while we spot misprints in eqs.~(184),(186),(187) of \cite{Hoang:2000fm},
we confirm correctness of eq.~(64) [apart from a `+' symbol
missing in the last line], 
which is the sum of eqs.~(183)--(187).
}
We list numerical values of the parameters for some representative
values of $x$, $\rho$ in Tab.~\ref{tab:param-mceff}.
\begin{table}
\begin{center}
\begin{tabular}{c|llllllll}
\hline
State & $\mu$~{\small [GeV]} & \,$\alpha_s(\mu)$ & ~~~$x$ & ~~$\rho$ & ~~$d_1^{(c)}$ & ~~$d_2^{(c)}$ & ~~~$\Delta _{\rm NLO}^{(c)}$ & ~~$\Delta _{\rm NNLO}^{(c)}$
\\
\hline
$\Upsilon (1S)$ & 
\small 
~~$5.352$ & \small $0.2092$ & \small $0.3050$ & \small $2.187$ & \small $0.4333$ & \small $11.66$ & \small $0.0008526$ & \small $0.002348$
\\
$\eta_b (1S)$ & 
\small 
~~$6.157$ & \small $0.2005$ & \small $0.3050$ & \small $2.282$ & \small $0.4333$ & \small $11.66$ & \small $0.0007655$ & \small $0.002113$
\\
\hline
\end{tabular}
\caption{\small
Numerical values of the parameters of non-zero charm mass effects,
for representative values of the input parameters.
We set $\overline m_b=4.18$~GeV and $\overline m_c=1.275$~GeV.
We evaluate
$\rho=3x/[2\alpha_s(\mu)]$ with the $\overline{\rm MS}$
masses; see explanation below eq.~(\ref{totb}).
}
\label{tab:param-mceff}
\end{center}
\end{table} 
There are no explicit spin-dependences in $d_{1,2}^{(c)}$ and
$\Delta _{\rm NLO,NNLO}^{(c)}$.
Their differences between $\Upsilon (1S)$ and $\eta_b (1S)$
originate only from the different values of $\mu$ chosen to evaluate
the energy levels.

After expressing $m^\text{pole}_{b,c}$ by $\overline m_{b,c}$ and applying
$\varepsilon$--expansion, we obtain
\begin{align}
&M_{b\bar b}=
\Bigl[ M_{b\bar b} \Bigr]_{\overline m_c \to 0}
+\varepsilon ^2 2\overline m_b \left\{ d_1^{(c)}\frac{\alpha_s^2}{\pi^2}-\Delta _{\rm NLO}^{(c)} \right\}
\nonumber \\
&~~~~~
+\varepsilon ^3 2\overline m_b \left\{
\left(d_2^{(c)} +2 d_1^{(c)}\cdot \frac{\beta_0}{2} {\rm log}\frac{\mu}{\overline m_b} \right)\frac{\alpha_s^3}{\pi^3}
-\frac{2 \alpha_s^4}{9\pi^2}d_1^{(c)}
-\Delta _{\rm NNLO}^{(c)}
-d_0\frac{\alpha_s}{\pi} \Delta _{\rm NLO}^{(c)}
\right\} ,
\label{totb}
\end{align}
where we show explicitly the
$\varepsilon$--expansion of the deviation from 
the ${\overline m_c\to 0}$ limit.
The term proportional to $\beta_0 \log (\mu /\overline m_b)$
arises from rewriting $\alpha_s(\overline m_b)$ by $\alpha_s(\mu)$;
the term proportional to $\alpha_s^4\, d_1^{(c)}$ from
the cross term of the leading-order binding energy and
$d_1^{(c)}$;
$m^\text{pole}_{b,c}$ in $\rho$
are replaced  by $\overline m_{b,c}$ without generating other terms
up to the order of our interest, since the ${\cal O}(\varepsilon)$ terms
cancel in the ratio $m^\text{pole}_c/m^\text{pole}_b$.

\begin{figure}[t]
\begin{center}
\includegraphics[width=9cm]{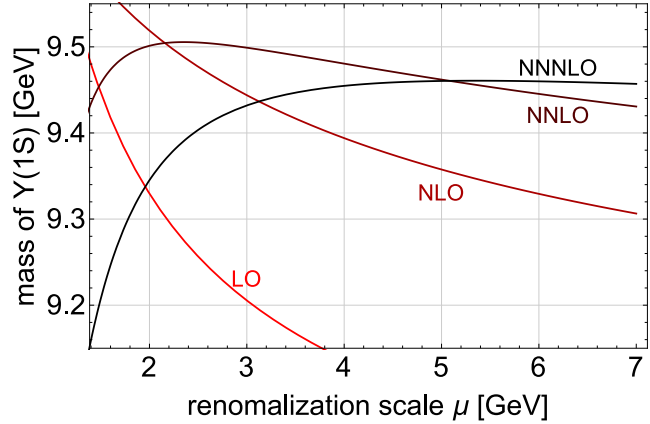}
\end{center}
\caption{\small
Scale dependences of the predictions for the $\Upsilon (1S)$ mass.
We take
$\overline m_b=4207$~MeV.
Other conventions are the same as in Fig.~\ref{fig:charmonium-scale-dep}.
}
\label{fig:bottomonium-scale-dep}
\end{figure}

We compare the predictions of $\Upsilon (1S)$ and $\eta_b (1S)$
masses with the experimental data \cite{Beringer:1900zz}:
\begin{align}
M_{\Upsilon (1S)}^\text{exp}=9460.30\pm 0.26 ~~ {\rm MeV},
~~~~~
M_{\eta_b (1S)}^\text{exp}=9398.0 \pm 3.2~~ {\rm MeV}\, .
\end{align}
The input $\alpha_s$ is taken as in eq.~(\ref{alpha}).
The $c$-quark mass in internal loops is taken as the PDG central value
$\overline m_c=1.275$~GeV in this analysis.\footnote{
Whether we vary $\overline m_c$ within the error $\pm25$~MeV or
choose the values in eqs.~(\ref{mcbar-Jpsi})--(\ref{mcbar-ave}),
variations of our predictions for $M_{b\bar b}$ are
much smaller and negligible compared to the uncertainties discussed below.
} 
We adjust the value of $\overline m_b$
to reproduce the experimental data.
The central values read, respectively, as
$\overline m_b=4.207$~GeV and 4.187~GeV.
The scale dependences of the predictions are shown
in Fig.~\ref{fig:bottomonium-scale-dep} for $\Upsilon (1S)$.
Those for $\eta_b (1S)$ are similar.
The minimal sensitivity scales are given by
$\mu =5.352$~GeV and 6.157~GeV, respectively.
The $\varepsilon$--expansions at the minimal sensitivity scales
are given by
\begin{align}
M_{\Upsilon (1S)}^\text{pert}&=8414 + 665+267+109+5~~ {\rm MeV}\, ,
\label{predicUpsilon}\\
M_{\eta_b (1S)}^\text{pert}&=8374 + 638+270+110+6~~ {\rm MeV}\, .
\label{predicetab}
\end{align}
We see reasonable stability and convergence of the predictions.

We examine separately the non-zero charm mass effects.
At each order of $\varepsilon$ in eq.~(\ref{totb}),
there is a  cancellation inside the curly bracket,
reflecting the cancellation of renormalons between
$2m^\text{pole}$ and $E_\text{bin}$.
The level of cancellation can be
quantified, e.g., by the ratio of 
the sum of the two terms at ${\cal O}(\varepsilon^2)$
and the sum of the absolute values of the two terms,
which is about 0.3--0.4 for 2~GeV$<\mu<6$~GeV.
The corresponding ratio at ${\cal O}(\varepsilon^3)$ is
about 0.1--0.2 for 2~GeV$<\mu<6$~GeV,
so that the cancellation is severer.

\begin{figure}[t]
\begin{center}
\vspace*{-5mm}
\includegraphics[width=7cm]{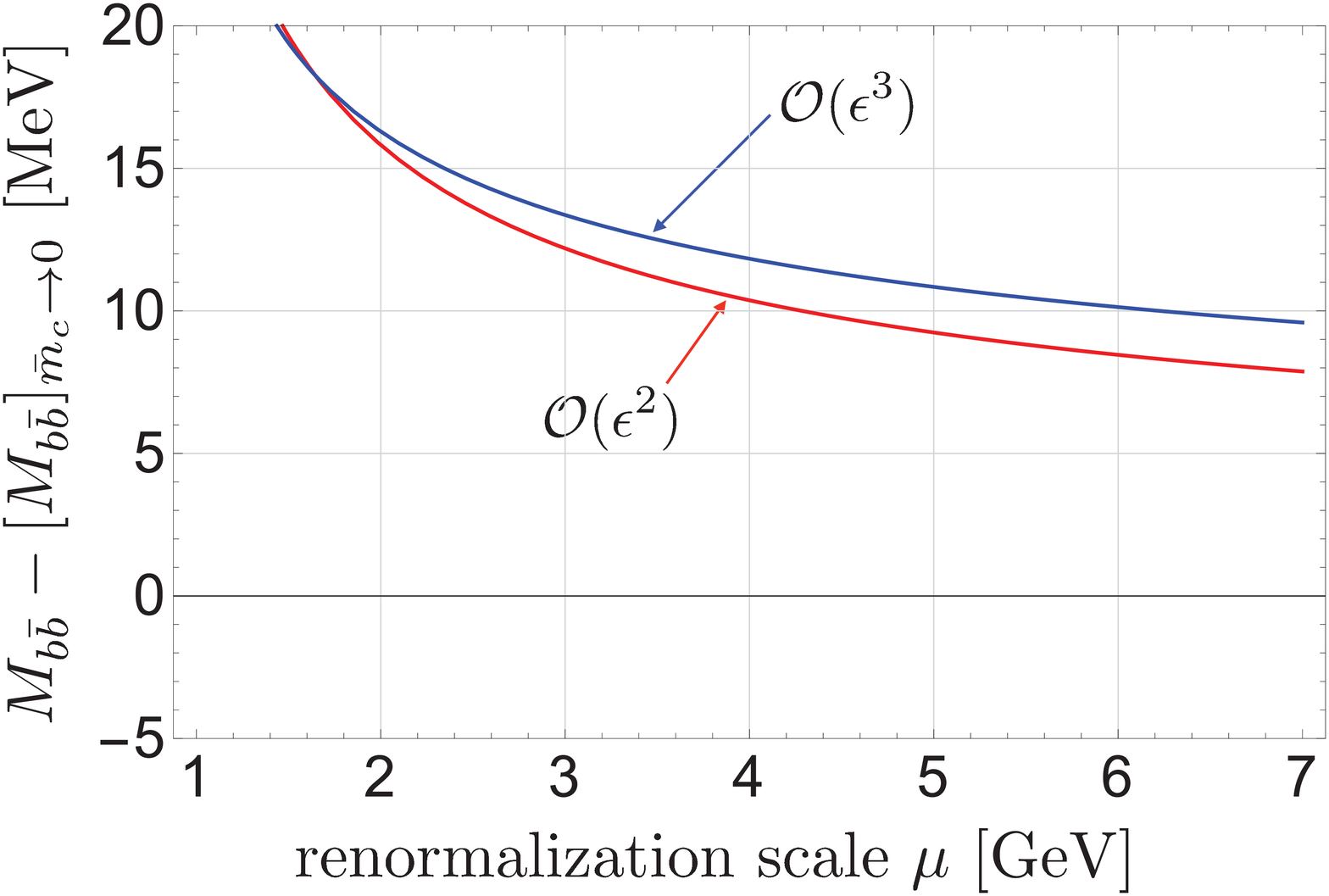}
~~~~
\includegraphics[width=7cm]{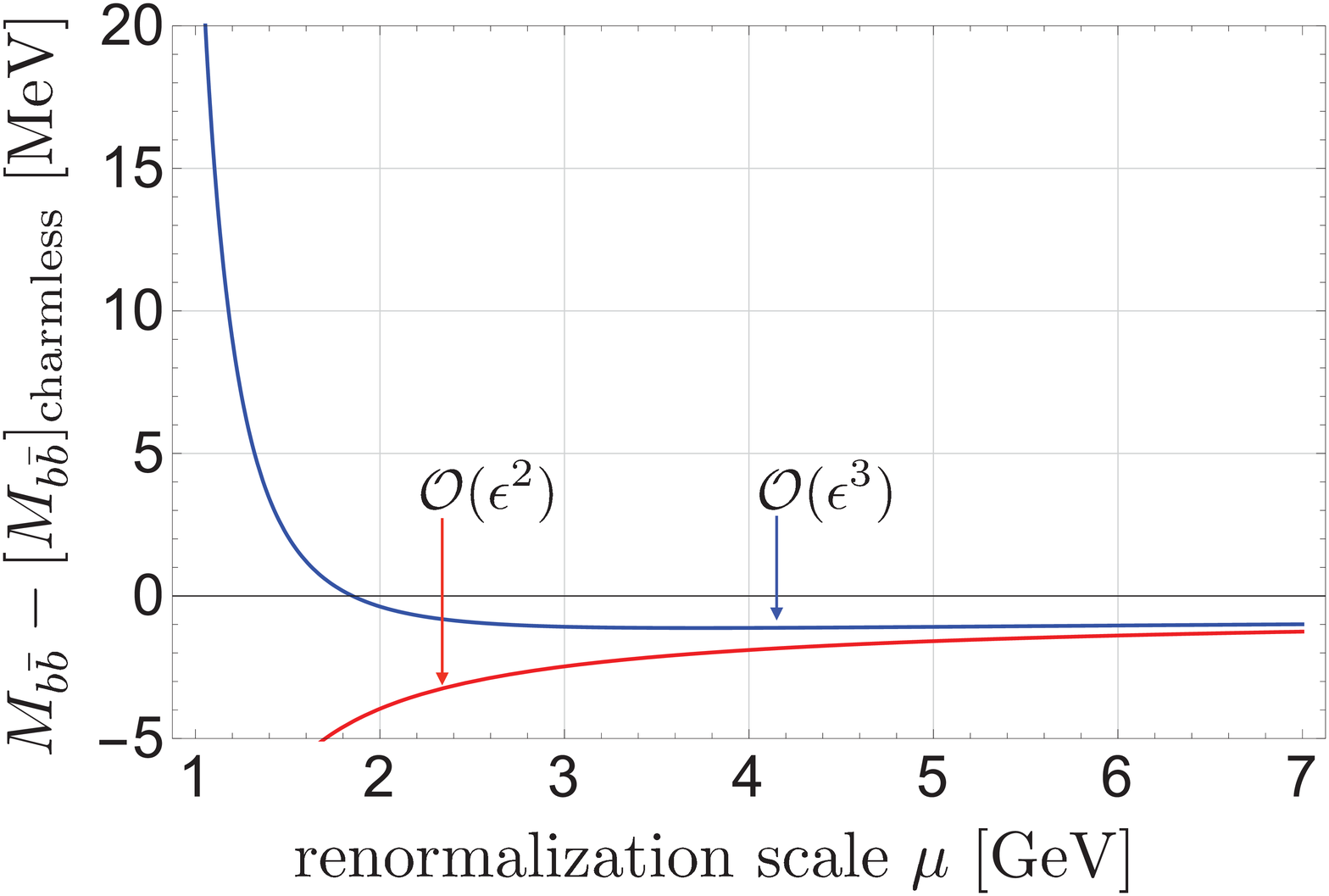}
\\
(a)\hspace*{70mm}(b)
\end{center}
\vspace*{-5mm}
\caption{\small
(a)
Scale dependences of the non-zero charm mass effects for
the bottomonium $1S$ states
at ${\cal O}(\varepsilon^2)$ and ${\cal O}(\varepsilon^3)$.
(b) Same as (a), but in the case that the energy level is
reexpressed in the 3-flavor coupling $\alpha_s^{(3)}$.
}
\label{fig:charm-mass-corr}
\end{figure}

Fig.~\ref{fig:charm-mass-corr}(a) shows the scale dependences of
the coefficients of $\varepsilon^2$ and $\varepsilon^3$ in
eq.~(\ref{totb}).
They are
comparable in size for $2~\text{GeV}\simlt \mu \simlt 6$~GeV.
Moreover, the scale dependence increases by including
the ${\cal O}(\varepsilon^3)$ term in addition to
the ${\cal O}(\varepsilon^2)$ term.
Hence, even though there are certain cancellations,
we do not observe convergence and stability of the
charm-mass correction by itself for the first two terms.\footnote{
Our result is also consistent with
the non-zero charm mass effects found
in the analyses using the non-relativistic sum rule \cite{Beneke:2014pta},
which chooses a scale $\mu\simeq 4.5$~GeV.
}
In refs.~\cite{Hoang:2000fm,Brambilla:2001qk}, 
an enhancement of the non-zero charm mass correction
was anticipated due to an accidental scale relation
$a_{1S}^{-1}\sim m_c \ll m_b$ ($a_{1S}$ is the
size of the bottomonium $1S$ states) and the resulting incomplete cancellation
of the ${\cal O}(\LQ)$ renormalons.
The above feature is consistent with this
expectation, which was put forward when the full 
${\cal O}(\varepsilon^3)$ term was still unknown.
In refs.~\cite{Brambilla:2001qk,Ayala:2014yxa}, an improvement in 
(apparent) convergence and stability is suggested by using
the 3-flavor coupling $\alpha_s^{(3)}$
instead of $\alpha_s^{(4)}$, since the renormalon-enhanced
effects are absorbed into the effective coupling $\alpha_s^{(3)}$.
Indeed we confirm the
improvement by this prescription.
See Fig.~\ref{fig:charm-mass-corr}(b), which shows the corresponding
${\cal O}(\varepsilon^2)$ and ${\cal O}(\varepsilon^3)$ terms
in the $\alpha_s^{(3)}$ scheme.
(We refer to \cite{Ayala:2014yxa} for the calculation method.)
Here, we 
use the 4-flavor coupling in our reference analysis, while
we use the 3-flavor coupling for an error estimate given below.
This is because
the $c$-quark cannot be regarded as completely decoupled at the
scale of the bottomonium $1S$ states \cite{Brambilla:2001qk}, and 
to use the effective 3-flavor coupling may not be natural.

By incorporating the charm-mass effects, qualitatively 
the potential energy $V(r)$
between $b$ and $\bar b$ becomes steeper
(interquark force becomes stronger) at large 
$r$ $(\simgt m_c^{-1})$, due to
the decoupling of the $c$-quark in the running of $\alpha_s$
\cite{Recksiegel:2001xq}.
This pushes up the energy levels of the bottomonium
for the same input $\overline m_b$.
As a result, the determined $\overline m_b$'s are reduced
compared to the $m_c\to 0$ case \cite{Kiyo:2013aea}
by about 8~MeV.

We estimate uncertainties in the determination of $\overline m_b$ in the
same way as in the charmonium case.
(The errors of the experimental data are negligibly small.)
\\
(i) Uncertainty due to uncertainty of $d_3$ is $\pm2$~MeV.
\\
(ii) Uncertainty due to uncertainty of $\alpha_s(M_Z)$ is
$\pm6$~MeV.
\\
(iii) Uncertainty by higher-order corrections.
(a) By changing $\mu$ to twice of the value of the
minimal sensitivity scale, $\overline m_b$ determined from
either $\Upsilon (1S)$ or $\eta_b (1S)$ varies by about 10~MeV.
(b) The differences in the determined $\overline m_b$ 
on the minimal sensitivity scales up to NNLO and NNNLO 
give 21~MeV and 18~MeV, respectively, for 
$\Upsilon (1S)$ and $\eta_b (1S)$.
(c) One half of the last known terms of eqs.~(\ref{predicUpsilon})
and (\ref{predicetab}) are both about 3~MeV.
We take as our estimates 
the maximal values of (a)--(c), namely, 
21~MeV and 18~MeV, respectively, for 
$\Upsilon (1S)$ and $\eta_b (1S)$.
Twice of these values are roughly of the same order of
magnitude as (or slightly larger than)
$\LQ^3 r^2$, in the case $\LQ\sim 300$~MeV and
$r \sim 1$~GeV$^{-1}$.

In addition, we estimate uncertainties of the
non-zero charm mass corrections.
\\
(iv) {\it Uncertainty of non-zero $\overline m_c$ effects}:
(a) The charm mass corrections at ${\cal O}(\varepsilon^2)$ and
${\cal O}(\varepsilon^3)$ shown in Fig.~\ref{fig:charm-mass-corr}(a)
are around 10~MeV at the minimal sensitivity scales 
$\mu\sim 5$--6~GeV of $M_{b\bar b}$.
We take the average of these two terms, which 
translates to about 5~MeV for the determined $\overline m_b$.
(b) We take the difference of the determined $\overline m_b$ by using
the 4-flavor coupling and the 3-flavor coupling.
This gives about 3~MeV  
for either $\Upsilon (1S)$ or $\eta_b (1S)$.
We take the maximal value of (a) and (b), namely 5~MeV, as our estimate.
This estimate of uncertainty from non-zero $\overline m_c$ 
is consistent with those of previous studies \cite{Penin:2014zaa,Beneke:2014pta}
using the non-relativistic sum rule.

\begin{figure}[t]
\begin{center}
\includegraphics[width=11cm]{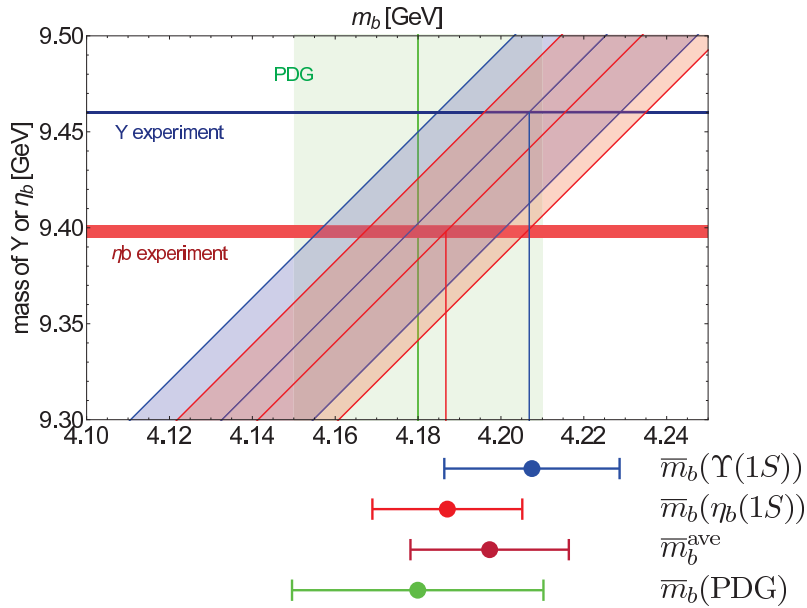}
\end{center}
\vspace*{-3mm}
\caption{\small
Determination of $\overline m_b$.
Horizontal (vertical) axis represents
$\overline m_b$ 
(mass of bottomonium $1S$ state).
Other notations are the same as in Fig.~\ref{fig:charm}.
}
\label{fig:bottom}
\end{figure}

To summarize our results, we obtain
\begin{align}
\overline m_b(\Upsilon (1S))&=4207\pm 2\ (d_3)\ \pm 6\ (\alpha_s)\ \pm 21\ (\text{h.o.} )  \pm 5 \ (m_c)\quad {\rm MeV} \, ,
\label{mbbar-Upsilon1}\\
\overline m_b(\eta _b (1S))&=4187\pm 2\ (d_3)\ \pm 6\ (\alpha_s)\ \pm 18\ (\text{h.o.} ) \pm 5 \ (m_c)\quad {\rm MeV}\, .
\label{mbbar-etab1}
\end{align}
Both values are mutually consistent within the estimated errors.
By taking the average of the above two estimates, we obtain
\begin{align}
\overline m_b^\text{ave}&=4197\pm 2\ (d_3)\ \pm 6\ (\alpha_s)\ \pm 20\ (\text{h.o.} )\pm 5 \ (m_c)\quad {\rm MeV}\, .
\label{mbbar-ave}
\end{align}
It is consistent with the current PDG
value $\overline m_b=4.18\pm0.03$~GeV.
(See Fig.~\ref{fig:bottom}.)

\subsection*{Conclusions and discussion}

We determined the $c$- and $b$-quark
$\overline{\rm MS}$ masses, by direct comparisons of the
experimental data for the masses of the individual heavy quarkonium $1S$
states with the
predictions of perturbative QCD.
The predictions combine the state-of-the-art computational
results, which are at the NNNLO level,
and show stability and convergence expected
for legitimate perturbative predictions.
The obtained values of each mass from the different
spin states
are consistent with each other
as well as with the current PDG value, which
is determined from a wide variety of observables.
The procedures of the computation and error estimates are  
based on fairly general prescriptions of
perturbative QCD.\footnote{
Calculation procedures were less systematized and unclear, 
when the same method was applied
more than a decade ago up to NNLO.
}

There is a general tendency that hyperfine splittings (also 
fine splittings) are predicted to be smaller than the
experimental data in perturbative
predictions of the heavy quarkonium energy levels.
This tendency is reflected in our analysis in
the differences of the determined values of
$\overline m_{b,c}$ from the different spin states.
There have been studies that particular higher-order corrections
increase the splittings
(hence, tend to remedy the differences), namely the corrections originating from
the running of $\alpha_s$ \cite{Recksiegel:2002za} and from
resummation of ultra-soft logarithms by the RG equation of potential-NRQCD
EFT \cite{Kniehl:2003ap}.
We did not include 
these corrections specific to the 
heavy quarkonium energy levels in our
estimates of higher-order corrections.
Rather we used more general methods applied in estimates
of unknown higher-order corrections
for various physical observables.
We remark that, in the end, we obtain consistent 
error estimates in both 
ways.

\section*{Acknowledgements}
The works of Y.K.\ and Y.S., respectively,
were supported in part by Grant-in-Aid for
scientific research Nos.\ 26400255 and 26400238 from
MEXT, Japan.
The work of G.M. is supported in part by Grant-in-Aid
 for JSPS Fellows (No.~26-10887).



\begin{thebibliography}{99}

\bibitem{Beringer:1900zz} 
  J.~Beringer {\it et al.} [Particle Data Group Collaboration],
  Phys.\ Rev.\ D {\bf 86}, 010001 (2012).

\bibitem{Penin:2014zaa} 
  A.~A.~Penin and N.~Zerf,
  JHEP {\bf 1404}, 120 (2014)

\bibitem{Beneke:2014pta} 
  M.~Beneke, A.~Maier, J.~Piclum and T.~Rauh,
  Nucl.\ Phys.\ B {\bf 891}, 42 (2015)

\bibitem{Dehnadi:2015fra} 
  B.~Dehnadi, A.~H.~Hoang and V.~Mateu,
  JHEP {\bf 1508}, 155 (2015)

\bibitem{Abramowicz:2014zub} 
  H.~Abramowicz {\it et al.} [ZEUS Collaboration],
  JHEP {\bf 1409}, 127 (2014)

\bibitem{Ayala:2014yxa} 
  C.~Ayala, G.~Cveti\v{c} and A.~Pineda,
  JHEP {\bf 1409}, 045 (2014).

\bibitem{Yang:2014sea} 
  Y.~B.~Yang {\it et al.},
  Phys.\ Rev.\ D {\bf 92}, no. 3, 034517 (2015)

\bibitem{Brambilla:2001fw}
  N.~Brambilla, Y.~Sumino and A.~Vairo,
  Phys.\ Lett.\ B {\bf 513}, 381 (2001).

\bibitem{Brambilla:2001qk} 
  N.~Brambilla, Y.~Sumino and A.~Vairo,
  Phys.\ Rev.\ D {\bf 65}, 034001 (2002).

\bibitem{Marquard:2015qpa} 
  P.~Marquard, A.~V.~Smirnov, V.~A.~Smirnov and M.~Steinhauser,
  Phys.\ Rev.\ Lett.\  {\bf 114}, 142002 (2015).

\bibitem{Hoang:2000fm} 
  A.~H.~Hoang,
  hep-ph/0008102.

\bibitem{Kniehl:2001ju}
  B.~A.~Kniehl, A.~A.~Penin, M.~Steinhauser and V.~A.~Smirnov,
  Phys.\ Rev.\ D {\bf 65} (2002) 091503


\bibitem{Kniehl:2002br}
  B.~A.~Kniehl, A.~A.~Penin, V.~A.~Smirnov and M.~Steinhauser,
  Nucl.\ Phys.\ B {\bf 635} (2002) 357.


\bibitem{Anzai:2009tm} 
  C.~Anzai, Y.~Kiyo and Y.~Sumino,
Phys.\ Rev.\ Lett.\  {\bf 104}, 112003 (2010);
  A.~V.~Smirnov, V.~A.~Smirnov and M.~Steinhauser,
Phys.\ Rev.\ Lett.\  {\bf 104}, 112002 (2010).

\bibitem{Kiyo:2014uca} 
  Y.~Kiyo and Y.~Sumino,
  Nucl.\ Phys.\ B {\bf 889}, 156 (2014)

\bibitem{Bekavac:2007tk} 
  S.~Bekavac, A.~Grozin, D.~Seidel and M.~Steinhauser,
  JHEP {\bf 0710}, 006 (2007).

\bibitem{Pineda:1997bj} 
  A.~Pineda and J.~Soto,
  Nucl.\ Phys.\ Proc.\ Suppl.\  {\bf 64}, 428 (1998);
  N.~Brambilla, A.~Pineda, J.~Soto and A.~Vairo,
  Nucl.\ Phys.\ B {\bf 566}, 275 (2000).
  
\bibitem{Luke:1999kz} 
  M.~E.~Luke, A.~V.~Manohar and I.~Z.~Rothstein,
  Phys.\ Rev.\ D {\bf 61}, 074025 (2000).

\bibitem{Sumino:2005cq}
  Y.~Sumino,
  Phys.\ Rev.\  D {\bf 76}, 114009 (2007).

\bibitem{Bauer:2011ws}
  C.~Bauer, G.~S.~Bali and A.~Pineda,
  Phys.\ Rev.\ Lett.\  {\bf 108}, 242002 (2012).

\bibitem{Sumino:2014qpa} 
  Y.~Sumino,
  {\it Lecture Note},
  arXiv:1411.7853 [hep-ph].

\bibitem{Kiyo:2013aea} 
  Y.~Kiyo and Y.~Sumino,
  Phys.\ Lett.\ B {\bf 730}, 76 (2014).

\bibitem{Hoang:1998ng} 
  A.~H.~Hoang, Z.~Ligeti and A.~V.~Manohar,
  Phys.\ Rev.\ Lett.\  {\bf 82}, 277 (1999).

\bibitem{Chetyrkin:1999qi} 
  K.~G.~Chetyrkin and M.~Steinhauser,
  Nucl.\ Phys.\ B {\bf 573}, 617 (2000)
  
\bibitem{Melnikov:2000qh} 
  K.~Melnikov and T.~v.~Ritbergen,
  Phys.\ Lett.\ B {\bf 482}, 99 (2000)
  
\bibitem{Chetyrkin:1997sg} 
  K.~G.~Chetyrkin, B.~A.~Kniehl and M.~Steinhauser,
  Phys.\ Rev.\ Lett.\  {\bf 79}, 2184 (1997)

\bibitem{Beneke:2005hg} 
  M.~Beneke, Y.~Kiyo and K.~Schuller,
Nucl.\ Phys.\ B {\bf 714}, 67 (2005).

\bibitem{Penin:2002zv}
  A.~A.~Penin and M.~Steinhauser,
  Phys.\ Lett.\ B {\bf 538} (2002) 335.

\bibitem{Gray:1990yh} 
  N.~Gray, D.~J.~Broadhurst, W.~Grafe and K.~Schilcher,
  Z.\ Phys.\ C {\bf 48}, 673 (1990).

\bibitem{Eiras:1999xx} 
  D.~Eiras and J.~Soto,
  Phys.\ Rev.\ D {\bf 61}, 114027 (2000)

\bibitem{Recksiegel:2001xq}
S.~Recksiegel and Y.~Sumino,
Phys.\ Rev.\ D {\bf 65}, 054018 (2002).

\bibitem{Recksiegel:2002za}
  S.~Recksiegel and Y.~Sumino,
  Phys.\ Rev.\ D {\bf 67} (2003) 014004;
  Phys.\ Lett.\ B {\bf 578} (2004) 369.

\bibitem{Kniehl:2003ap}
  B.~A.~Kniehl {\it et al.},
  Phys.\ Rev.\ Lett.\  {\bf 92} (2004) 242001
   [Erratum-ibid.\  {\bf 104} (2010) 199901].









  
\end{thebibliography}
\end{document}